\documentclass[12pt,amsfonts]{article}

\def\vp{\varphi}

\def\half{\textstyle{\frac{1}{2}}}

\def\half{\textstyle{\frac{1}{2}}}

\def\H{{\cal H}}

\def\th{\theta}
\def\de{\delta}
\def\H{{\cal H}}
\def\G{{\cal{G}}}

\def\D{{\cal{D}}}

\def\N{{\cal{N}}}

\def\ra{\rightarrow}
\def\tint{{\textstyle\int}}
\def\hg{{\hat g}}
\def\hp{{\hat\pi}}

\def\s{\hskip.08em}

\def\d{\partial}

\def\b{\begin{eqnarray*}}  %takes no eqn numbers
\def\e{\end{eqnarray*}}    %takes no eqn numbers
\def\bn{\begin{eqnarray}}  %takes eqn numbers
\def\en{\end{eqnarray}}   %takes eqn numbers

\def\<{\langle}

\def\>{\rangle}

\def\no{\nonumber}

\def\pie{\hat{\pi}}
\def\gg{\hat{g}}

\def\{{\lbrace}
\def\}{\rbrace}
\def\W{{\mathcal{W}}}
\def\om{\omega}
\def\Op{{\cal{O}}}

\begin{document}  

\title{The Unification of \\Classical and Quantum Gravity}        
\author{John R. Klauder\footnote{klauder@phys.ufl.edu} \\
Department of Physics and Department of Mathematics \\
University of Florida,   %P.O. Box 118440\\
Gainesville, FL 32611-8440}
\date{ }
%let\frak\cal
\bibliographystyle{unsrt}
%\begin{document}
\maketitle
\begin{abstract} The favored classical variables that are promoted to quantum operators are divided into three sets that feature constant positive curvatures, constant zero curvatures, as well as constant negative curvatures. This list covers the spin variables, the canonical variables, and the affine variables, and these three topics will be briefly reviewed. In this discussion, appropriate coherent states are introduced which are the principal items that are critical in the unification of relevant classical and quantum realms. This analysis can also serve to unify classical gravity and quantum gravity without any speculative aspects.
\end{abstract}
\section{Introduction}
The rules governing favored classical variables and their sets of favored quantum operators lie in three separate divisions, namely, canonical variables, spin variables, and affine variables.
This section will introduce the first two sets briefly. and feature the third set more completely because that set is relevant to the primary topic of this paper, i.e., affine classical and quantum variables, that arise in our study.

\subsection{Canonical theory analysis}
The favored classical variables in this section, $p$ and $q$, have the properties that $-\infty < p,q<\infty$, and a Poisson bracket given by $\{q,p\}=1$. A pair of quantum operators, $P$ and $Q$,
which obey $[Q,P]=i\hbar1\!\!1$ are paired to create canonical coherent states given by
   $|p,q\>\equiv e^{-iqP/\hbar}\,e^{ipQ/\hbar}\,|\omega\>$, where the fiducial unit vector $|\omega\>$ obeys $(Q+iP/\omega)|\omega\>=0$, which implies that $\<\omega| Q|\omega\>=\<\omega|P|\omega\>=0$. We choose $Q$ as dimensionless, and $P$ and $\om$ having the dimensions of $\hbar$.
   For a general operator, it follows that $W(p,q)\equiv\<p,q|\W(P,Q)|p,q\>=\<\omega|\W(P+p,Q+q)|\omega\>=\W(p,q)+\Op(\hbar;p,q)$.
If $\hbar\ra0$ then $\Op\ra0$, and the quantum variables in $\W$ occupy the same position as the classical variables in $W$, which Dirac proposed \cite{dirac} as part of the rules to identify the favored variables. The final rule of Dirac was that the favored classical variables should be Cartesian coordinates, although he did not prove this last requirement. In an analysis that features the coherent states in rays (i.e., independent of any real, operator free, phase factor), it follows \cite{a,2} that $d\sigma(p,q)^2\equiv 2\hbar[\,|\!|\,d|p,q\>\,|\!|^2-|\<p,q|\,d|p,q\>|^2\,]=\om^{-1}\,dp^2+ \om\,dq^2$, establishing that these variables are indeed the favored set. We emphasize that the
favored variables here exist on a `constant zero curvature', which is a two-dimensional flat space.

These canonical coherent states possess a completeness that spans the entire Hilbert space as
$\int |p,q\>\<p,q|\; dp\, dq/ 2\pi\hbar=1\!\!1$, and their over completeness leads to $\<p,q|\W|p,q\>=0$ for all $p,q$ implies that $\<p,q|\W|p',q'\>=0$ for all $p,q,p',q'$, i.e. $\W=0$ \cite{c}.
This emerges  from the analysis that there is some expression $w(r,s)$ that leads to $\W=\tint
w(r,s)\,|r,s\>\<r,s|\; dr\,ds/2\pi\hbar$, which therefore leads to 
%w$\<p,q|\W|p,q\>=\tint w(r,s)\, |\<p,q|r,s\>|^2\;dr \,ds/2\pi\hbar=\tint\,w(r,s)\, e^{􏰀-[\om^{-1}(r-%p)^2+\om(s-q)^2]/2\hbar} \;dr\,ds/2\pi\hbar$. 
$\<p,q|\W|p,q\>=\tint w(r,s)\, |\<p,q|r,s\>|^2\;dr \,ds/2\pi\hbar=\tint\,w(r,s)\, e^{-[\om^{-1}(r-p)^2+\om(s-q)^2]/2\hbar} \;dr\,ds/2\pi\hbar$. 
Hence, if  $\<p,q|\W|p,q\>=0$ for all $p,q$, then $\W=0$. 

This last property may not hold for eigenfunctions. For example, the set of vectors, $|x\>$, where $Q|x\>=x|x\>$, lead, with $c\neq 0$, to $\<x|e^{icP}|x\>=0$ for all $x$, but $\<x|e^{icP}|x'\>\neq 0$ when $x'=x-c$.

The principal purpose of this paper is to offer a unification of classical realms and quantum realms 
with smooth and natural procedures. The author's paper \cite{b} outlined such  procedures for
the three realms of classical analysis, but particularity featured a `bridge' which unifies the 
canonical classical and quantum realms. For readers interested in that story the author's previous 
paper \cite{b} can be recommended. The unification of classical and quantum gravity will be treated in Sec. 2 of the present paper.
 
\subsection{Spin theory analysis}
The classical variables $-\pi/2\leq\th\leq\pi/2$ and $-\pi<\vp\leq\pi$ are paired with three spin operators
$S_1,S_2,S_3$ which obey $[S_i,S_j]=\hbar\,\epsilon_{i,j,k}\,S_k$, and $\sum_j S^2_j=\hbar^2s(s+1)1\!\!1_s$, (with $Tr(1\!\!1_s)=2s+1)$, where, for $SU(2)$ and $SO(3)$, $2s\in\{1,2,3,...\}$. The operator $S_3$ is diagonalized so that $S_3|s,m\>=m\hbar|s,m\>$, where $m\in\{-s,...s-1,s\}$, and 
the spin coherent states are $|\th,\vp\>\equiv e^{-i\vp S_3/\hbar} e^{-i\th S_2/\hbar}\,|s,s\>$.
The expectations $\<s,m|S_j|s,m\>$ are $0$ for $j=1,2$, for all $m$, while it is $m\hbar$ for $j=3$.
The analog of Cartesian coordinates leads to $d\sigma(\th,\vp)^2\equiv 2\hbar [\,|\!|\,d|\th,\vp\>\,
|\!|^2-|\<\th,\vp|\,d|\th,\vp\>|^2\,]=\hbar\,s [\,d\th^2+ \cos(\th)^2\,d\vp^2]$, which describes the surface of a three-sphere with a radius of $(\hbar\,s)^{1/2}$. In particular, the surface here is that of a `constant positive curvature' with a magnitude of $ (\hbar s)^{-1}$.

The spin coherent states possess a completeness that spans their Hilbert space as
$\tint |\th,\vp\>\<\th,\vp| \,(2s+1) \cos(\th)\, d\th\,d\vp/4\pi=1\!\!1_s$ \cite{c}. Moreover, if $\G$ is composed of the three spin operators, and if $\<\th,\vp|\G|\th,\vp\>=0 $ for all $\th, \vp$, then $\G=0$.

The unification of the spin classical realm and the spin quantum realm was 
sufficiently explained in \cite{b}, and therefore it will not be repeated again in the present paper.

\subsection{Affine theory analysis}
Affine quantization is similar to canonical quantization, but they both work well for separate sets of problems. Even for a half-harmonic oscillator, where $0<q<\infty$, canonical quantization fails and affine quantization succeeds \cite{2,d}. With the limited coordinate $q>0$, the momentum operator $P$ can not be made self adjoint. A new classical variable, $d\equiv pq$, takes the place of $p$, and so $d$ and $q$ are the new classical variables promoted to the operators $D\equiv(PQ+QP)/2$ and $Q>0$, which obeys  $[Q,D]=i\hbar\, Q$, and observe that we nevertheless will still use $p$ and $q>0$ as classical variables.
The affine coherent states are given by (note that ; replaces ,)  $|p;q\>\equiv
e^{ipQ/\hbar} e^{-i\ln(q)D/\hbar}\,|\beta\>$, in which we chose $q$ and $Q$ as dimensionless, and thus
$p$, $D$, and $\beta$ have the dimensions of $\hbar$. The fiducial vector is chosen so that 
$[(Q-1\!\!1)+iD/\beta]|\beta\>=0$, which implies that $\<\beta|Q|\beta\>=1$ and $\<\beta|D|\beta\>=0$. The classical variables $p;q$ can not be Cartesian because $q>0$, and we again choose the special metric \cite{a} to find that $ d\sigma(p;q)^2\equiv 2\hbar[\,
|\!|\, d|p;q\>\,|\!|^2-|\<p;q|\,d|p;q\>|^2\,]=\beta^{-1}q^2\,dp^2+ \beta\,q^{-2}\,d q^2$,
which is certainly not Cartesian. However, this specific metric has a `constant negative curvature' with a value of $-2/\beta$, a property that is not visible in our 3-dimensional spatial dimensions \cite{f}.

The affine coherent states admit a resolution of the identity \cite{c} given by the relation
$\tint |p;q\>\<p;q| \,[1-\hbar/2\beta]\; dp\,dq/4\pi\hbar=1\!\!1$, provided that $\beta>\hbar/2$.
Once again we find that if $\<p;q|\W|p;q\>=0$ for all $p;q$, it follows that $\W=0$. The proof of this claim assumes that $w(p;q)$ exists such that $\W=\tint \,w(p;q)\,|p;q\>\<p;q|\,[1-\hbar/2\beta]\;
dp\,dq/4\pi\hbar$, again provided that $\beta>\hbar/2$. Thus $\<r;s|\W|r;s\>=\tint\,w(p;q)\,|\<r;s|p;q\>|^2
\,[1-\hbar/2\beta]\; dp\, dq/4\pi\hbar=\tint\,w(p;q)\,(sq)^{-2\beta/\hbar}/\{[(1/s+1/q)^2+\beta^{-2}(r-p)^2
]/4\}^{2\beta/\hbar}\,[1-2\hbar/\beta]\;dp\,dq/4\pi\hbar$; this is all with $s>0$ and $q>0$. Again, we see that if $\<p;q|\W|p;q\>=0$ for all $p;q$, then $\W=0$. 

This strength of a pair of identical coherent states used to generate the expectation value of an arbitrary operator will be drawn on in our unification of classical and quantum gravity.

\subsubsection{Unification of affine classical and quantum stories}
Since affine quantization will be the theory used to unify classical and quantum gravity, we choose
to offer a miniature version of how that effort will appear with gravity. We begin our toy example 
with the introduction of the quantum action functional given by
  \bn A_{q}=\tint_0^T\{ \<\psi(t)|[i\hbar(\d/\d t)-\H'(D,Q)]|\psi(t)\>\}\;dt\;, \label{wqw}\en
and stationary variations lead to a form of Schr\"odinger's equation
   \bn i\hbar (\d\,|\psi(t)\>/\d t) =\H'(D,Q)\,|\psi(t)\>\;.\en
   
Next, we introduce a semi-classical action functional for this story given by
    \bn &&\hskip-2emA_{sc} =\tint_0^T\{\<p(t);q(t)|[i\hbar(\d/\d t)-\H'(D,Q)]|p(t);q(t)\>\}\;dt \no \\
     &&\hskip-.2em=\tint_0^T \{\<\beta|[-\dot{p}(t)q(t)Q+\dot{q}(t)D/q(t)-\H'(D+p(t)q(t)Q,q(t)Q)]|
     \beta\>\}\;dt \;.\no \\ \label{yy}
     &&\hskip-.2em= \tint_0^T \{-\dot{p}(t)q(t)-H'(p(t)q(t), q(t))\} \;dt \;, \label{98}\en
     where $H'(pq,q)=\H'(pq,q)+\Op(\hbar;p,q)$. If $\hbar\ra0$ then $H'(pq,q=\H'(pq,q)$, or
     $\Op\hbar;p,q)$ may be so small it can be ignored; in that case the last line in (\ref{yy}) is 
     considered to be the classical equation. 
     
     The unification of these realms is secured by a mathematical `bridge' 
     \bn B(r;s|r';s')\equiv\<r;s|[i\hbar(\d/\d t)-\H'(D,Q)]|r';s'\> \en
     which leads to the integrand for the 
     classical action functional or the quantum action functional if the `bridge'\footnote{Figure 1 in \cite{b} is a pictorial representation of a `bridge' permitting a smooth and continuous connection between classical and quantum realms.} is coupled to 
     two resolutions of the identity, where $C\equiv[1-\hbar/2\beta]/4\pi\hbar>0$,
     with
  \bn C^2\!\int\!\int \<p(t);q(t)|r;s\>\,B(r;s|r';s')\,\<r';s'|p(t);q(t)\> \;dr\,ds\;dr'\,ds' \en
  to lead to the classical action functional integrand (\ref{98}), or two resolutions of the identity with       
  \bn C^2\!\int\!\int \<\psi(t)|r;s\>\,B(r;s|r';s')\,\<r';s'|\psi(t)\>\;dr\,ds\;dr'\,ds' \en
     to lead to the quantum action functional integrand (\ref{wqw}), with integrals in both cases over the pair $r$; $s$ as well as the pair $r'$; $s'$. 
     
     Either of these procedures creats a genuine affine action integrand for either the classical action functional or the quantum action functional, demonstrating a smooth and continuous connection between the classical and quantum realms. 
     
\section{Unification of Classical and \\Quantum Gravity}
The affine quantization approach that has been chosen to deal with gravity, and which the author has used in recent publications dealing with gravity \cite{2,c,b,j2,j3,111,j4,j5}, will experience a unification unlike any other approach toward classical and quantum gravity. The classical formulation of gravity that is chosen  considers a spacial slice that undergoes a temporal advancement. A procedure such at that, known as the ADM approach \cite{78}, consists of essential variables as well as additional variables that need to be eliminated through the presence of various constraints. While constraints are handled well enough classically, quantizations involving constraints may cause some problems. One such constraint is that the Hamiltonian density should vanish, and as a quantum constraint that the Hamiltonian operator is limited to Hilbert space vectors that have a subset of eigenvalues that vanish. To do so properly, it is necessary that the Hamiltonian operator is well defined prior to restricting its physically important spectrum. Canonical quantization efforts to quantize gravity have encountered 
difficulties in ensuring a proper operator, and, fortunately, affine quantization is successful in this issue as will be observed below. Let us now begin our approach to create a smooth and continuous unification of the classical and quantum realms.

\subsection{Classical gravity according to ADM}
The set of classical variables includes the metric $g_{ab}(x)\;(=g_{ba}(x))$ and the momentum $\pi^{cd}(x)\;(=\pi^{dc}(x))$, where $ a,b,c,d,...=1,2,3$. The metric is positive definite, i.e., $g_{ab}(x)\,dx^a\,dx^b>0$, which makes the determinant, $g(x)\equiv\det[g_{ab}(x)]>0$, as well. 
We also introduce the momentric $\pi^q_b(x)\equiv \pi^{ac}(x)\,g_{bc}(x)$.\footnote{The name momentric is a mix of {\it momen}tum and me{\it tric}.} The classical action functional is given by
 \bn &&\hskip-3em A_c=\tint_0^T\tint\{ -g_{ab}(x,t)\,\dot{\pi}^{ab}(x,t) -N^a(x,t)H_a(x,t) -N(x,t) H( x,t)\}\: d^3\!x\,dt, \en
    where $N^a(x,t)$ and $N(x,t)$ are Lagrange multipliers that enforce the vanishing of
the diffeomorphism constraint $H_a(x,t)=\pi^b_{a|b}(x,t)$, where $_|$ denotes a covariant derivative, and the Hamiltonian constraint is
   \bn &&H(x,t)=g(x,t)^{-1/2}[\pi^a_b(x,t)\pi^b_a(x,t)-\!\half 
   \pi^a_a(x,t)\pi^b_b(x,t)] \no \\ &&\hskip12em+g(x,t)^{1/2}R(x,t)\;, \en
   where $R(x,t)$ is the three-dimensional scalar curvature.
   
 \subsection{Affine quantization of ADM gravity}
 Canonical quantization of gravity promotes the metric and the momentum fields to quantum operators. However, the positivity restrictions on the metric mean that the momentum operators can not be self adjoint. Instead, affine quantization promotes the metric and momentric fields, and both operator sets can be self adjoint while metric positivity is preserved. Promoted from Poisson brackets, these operators have the following commutations  
 \bn   &&[\hp^a_b(x),\s \hp^c_d(x')]=i\s\half\,\hbar\,\delta^3(x,x')\s[\delta^a_d\s \hp^c_b(x)-\delta^c_b\s \hp^a_d(x)\s]\;,    \no \\
       &&\hskip-.10em[\hg_{ab}(x), \s \hp^c_d(x')]= i\s\half\,\hbar\,\delta^3(x,x')\s [\delta^c_a \hg_{bd}(x)+\delta^c_b \hg_{ad}(x)\s] \;, \label{39} \\
       &&\hskip-.20em[\hg_{ab}(x),\s \hg_{cd}(x')] =0 \;. \no  \en
 It follows that the Hamiltonian operator, $\H'(\pie^a_b,\gg_{cd})$, can also be self 
 adjoint, as was the goal.

    \subsection{Affine coherent states for gravity}
    We choose the metric and momentric operators to build our coherent states for gravity, specifcally,\footnote{Some terms are given a different notation than those used in \cite{2}.}
   \bn |\pi; g\>\equiv e^{(i/\hbar)\tint \pi^{ab}(x)\hat{g}_{ab}(x)\; d^3\!x }\; e^{-(i/\hbar)\tint\eta^a_b(x)\hat{\pi}^b_a(x)\;d^3\!x}\;|b\>\;, \en
   where $\pi$ stands for $\{\pi^{ab}\}$ and $g$ stands for $\{g_{cd}\}$.
   The fiducial vector $|b\>$ satisfies the relation $[(\gg_{ab}(x)-\delta_{ab}1\!\!1)+
   i\pie^c_d(x)/b(x)\hbar]|b\>=0$, which implies that $\<b|\gg_{ab}(x)|b\>=\delta_{ab}$ as well as
   $\<b|\pie^c_d(x)|b\>=0$.
   The choice of $|b\>$ is such 
   that the matrix $\eta(x)\equiv\{\eta^a_b(x)\}$ enters the coherent states solely in the form given by
          \bn \langle\pi;g|\hat{g}_{ab}(x)|\pi; g\rangle =[e^{\eta(x)/2}]^c_a\,\langle b|\hat{g}_{cd}(x)\,|b\rangle\,[e^{\eta(x)/2}]^d_b\equiv g_{ab}(x)\;, \label{mm} \en
          which ensures that $\{g_{ab}(x)\}>0$ as required,\footnote{Any real, $3\times 3$ matrix, like  $\eta$, that is symmetric, i.e., $\eta^T=\eta$, and which has been exponentiated, i.e., $e^{\eta}$, becomes a symmetric, positive definite matrix that can serve to become our metric. The exponential of $\eta$ is a gift from the basic affine operators.} and this expression helps clarify the naming of the gravity coherent states. A companion relation is given by
    \bn \langle\pi, g|\hat{\pi}^a_b(x)|\pi, g\rangle
=\pi^{ac}(x)\,g_{cb}(x)\equiv\pi^a_b(x)\;,\en
which also involves the metric result from (\ref{mm}). 
               As a consequence, the inner product of two gravity coherent states is given by 
   \bn \langle\pi'',g''|\pi',g'\rangle\hskip-1.3em&&=\exp\Big{\{}\textstyle{-2\int}b(x)\,d^3x \\
   &&\hskip-3em \times\ln\big\{
\frac{\det\{\frac{1}{2}[ {g''}^{ab}(x)+{g'}^{ab}(x)]+i\frac{1}{2\hbar}b(x)^{-1}[{\pi''}^{ab}(x)-{\pi'}^{ab}(x)]\}}{\det[{g''}^{ab}(x)]^{1/2}
\,\,\det[{g'}^{ab}(x)]^{1/2}}  \big\} \Big\}\;, \no \en
where the scalar density function $b(x)>0$ ensures the covariance of this expression.

To test whether or not we have `favorable coordinates' we examine, with a suitable factor $J$, the Fubini-Study metric \cite{a} given by
 \bn d\sigma(\pi;g)^2\hskip-1.4em&&\equiv J\hbar[\,\|\,d|\pi;g\>\|^2-|\<\pi;g|\;d|\pi;g\>|^2\,] \no \\
     &&=\tint \{(b(x)\hbar)^{-1} g_{ab}(x)\,g_{cd}(x)\,d\pi^{bc}(x)\,d\pi^{da}(x) \\
     &&\hskip2em +(b(x)\hbar) \,g^{ab}(x)\,g^{cd}(x)\,dg_{bc}(x)\,dg_{da}(x)\}\; d^3\!x \;.\no \en
    This metric represents a multiple family of constant negative curvature spaces. Based on the previous analysis, we accept that
    the basic affine quantum variables have been promoted from basic affine classical variables.
    
    Much like the resolution of the identity for the coherent states $|p;q\>$ in Sec.~1.3, we are able to obtain a resolution of unity for the gravity coherent states, namely by
      \bn 1\!\!1=\N\int |\pi;g\>\<\pi;g| \;\D \pi\,\D g\;, \en
      for a suitable constant $\N$. One way to determine $\N$ is to ensure that
      \bn 1=\<\pi';g'|1\!\!1|\pi';g'\>=\N\int\ |\<\pi';g'|\pi;g\>|^2 \;\D\pi\,\D g\;. \en
      Regularizing the continuum of $x$ into a lattice, followed by a reduction back to the continuum,
      is also a good way to calculate $\N$.

\subsection{The semi-classical affine gravity action functional}
The classical Hamiltonian density suggests the form of the quantum Hamiltonian as given by
  \bn  &&\hskip-4em \H'(\pie^a_b,\gg_{cd})=\pie^a_b(x)\gg^{-1/2}(x)\pie^b_a(x)-\half \pie^a_a(x)\gg^{-1/2}(x)\pie^b_b(x)\no \\ &&\hskip7em+\gg(x)^{1/2}\,\hat{R}(x)\;, \en which leads us to the semi-classical gravity action functional given by
   \bn &&A_{sc}=\tint_0^T\tint \{\<\pi;g|[i\hbar(\d/\d t)-\H'(\pie^a_b,\gg_{cd})]|\pi;g\>\}\;d^3\!x\,dt \no \\
        &&\hskip1.8em=\tint_0^T\tint\{\<b|[-\dot{\pi}^{ab}
        [e^{\eta/2}]_a^c \gg_{cd}[e^{\eta/2}]^d_b 
         +\dot{\eta}^a_b[e^{\eta}]^b_c\pie^c_a \no \\
         &&\hskip4em-\H'(\pie^a_b+\pi^{ac}[e^{\eta/2}]_c^e \gg_{ed}[e^{\eta/2}]^d_b,
          [e^{\eta/2}]_c^a\gg_{ab}[e^{\eta/2}]^b_d) ]|b\> \}\; d^3\!x\,dt \no\\
         &&\hskip2em=\tint_0^T\tint \{ -\dot\pi^{ab}(x,t)g_{ab}(x,t)-H'(\pi^a_b(x,t),g_{cd}(x,t))\}
         \;d^3\!x\,dt \;, \label{00} \en
   and it follows that $H'(\pi^a_b,g_{cd})=\H'(\pi^a_b,g_{cd})+\Op(\hbar; \pi,g)$, leading to the 
   usual classical limit as $\hbar\ra0$ or $\Op$ is tiny enough to ignore.
   
\section{The Unification of Classical and Quantum Gravity}
To begin this section we introduce the quantum affine action functional, including only the Hamiltonian, which is given by
  \bn A_q=\tint_0^T\tint \{\<\Psi(t)|[i\hbar(\d/\d t)-\H'(\pie^a_b(x), \gg_{cd}(x))]|\Psi(t)\>
  \}\;d^3\!x\, dt \;.\label{yes}\en
  
We next introduce a mathematical expression that we call the `bridge', and which is given by 
\bn B(\pi;g|\pi';g')\equiv \tint\<\pi;g|[i\hbar(\d/\d t)-\H'(\pie^a_b(x),\gg_{cd}(x))]|\pi';g'\>\;d^3\!x \;. \label{hh}\en
   This expression, despite its close appearance to a semi-classical integrand, has no role in
   the physical expressions that belong to the classical or the quantum realms.
  Although we can restore the semi-classical functional integrand simply by setting 
  $|\pi';g'\>\ra|\pi;g\>$, we instead choose another path to seek the classical realm, and we let 
  the `bridge' take us to the expression
   \bn \N^2\!\int\!\int\{\<\pi(t); g(t)|\pi;g\> \,B(\pi;g|\pi';g')\<\pi';g'|\pi(t);g(t)\>\}\; \D\pi\,\D g\;\;\D\pi'\,D g' \;, \en
   which determines the integrand for the physical classical action functional (\ref{00}),
   and is ready to provide any contribution needed at this point. 
   
   Finally, we let the `bridge' take us to the quantum realm by the expression 
    \bn \N^2\!\int\!\int\{\<\Psi(t)|\pi;g\> \,B(\pi;g|\pi';g')\<\pi';g'|\Psi(t)\>\}\; \D\pi\, \D g\;\;\*D\pi'\,D g' \;, \en
    which determines the integrand for the physical quantum action functional (\ref{yes}),
    and is ready to provide any contribution needed at this point.
    
    The relation in (\ref{hh}) requires more information than the same expression with $|\pi';g'\>=
    |\pi;g\>$, as if $\<\pi;g|\pi;g\>=1$, you can not learn what is $\<\pi;g|\pi';g'\>$. Even knowing
    $\<\Psi|\W|\Psi\>$ for all states does not tell you what $\<\pi;g|\W|\pi';g'\>$ is since, for a self-adjoint $\W$, the diagonal term is real while the non-diagonal term can be complex. This points to the fact that the `bridge' holds information that is outside the classical or quantum realm. 
    Happily, the `bridge' offers us a smooth and continuous path between the classical and quantum realms. 
    
    It follows that the `bridge' has introduced
    a genuine path that unifies the classical and quantum realms, as
    promised by the title of this paper.
    
    \subsubsection{Additional items}
    
    {\it Diffeomorphism constraint}
    
    This constraint, when quantized, is $\pie^a_{b|a}(x)$, and in the semi-classical form it becomes
 \bn &&\<\pi;g|\,\pie^a_{b|a}(x)\,|\pi:g\>=\<b| \{\pi(x)^{ac}(x)
 [e^{\eta(x)/2}]_c^d \gg_{de}(x))[e^{\eta(x)/2}]^e_b\}_{|a}|b\>\no \\
 &&\hskip11em =\{\pi^{ac}(x)g_{cb}(x)\}_{|a} =\pi^a_{b|a}(x)\;. \en
 {\it Lagrange  multipliers}
 
 The Lagrange multipliers, $N(x)$ and $N^b(x)$ remain $c$-numbers in the quantization and enforce the facts that $\H'(\pie^a_b(x),\gg_{cd}(x))$, and
 $\pie^a_{b|a}(x)$ both require a reduction in Hilbert space vectors in which all of their eigenvectors involve vanishing eigenvalues.\vskip,8em \hskip-1.6em
 {\it Schr\"odinger's representation}
 
 The equations offered in (\ref{39}) permit $\gg_{ab}(x)\ra g_{ab}(x)$ and 
  \bn \pie^a_b(x)\ra -\half i\hbar[ g_{bc}(x)\,(\de/\de g_{ac}(x))+(\de/\de g_{ac}(x))g_{bc}(x)] \;.\en
For the Schr\"odinger equation, see Eq.~(46) in \cite{2}. Regularization of the spatial variable
$x$ as a discrete, three-dimensional lattice proves useful, which 
appears, e.g., in \cite{v}.
    
   \section{Remaining Issues to Complete the \\Quantization of Gravity}
   This paper has been devoted to ensuring a proper Hamiltonian operator for the quantization of 
   gravity by following similar procedures developed in \cite{b}. Even though the Hamiltonian constraint retains only the eigenfunctions that are focused on zero eigenvalues, it is necessary to establish that the Hamiltonian operator be well defined. The
   Hamiltonian constraint, as well as the diffeomorphism constraint, 
   can both be 
   incorporated into a general constraint procedure \cite{v,mm} that overcomes the usual difficulties
   found when dealing with constraints, such as second-class constraints. In addition, the paper \cite{v} also introduces a Thiemann-like master constraint expression \cite{ss}. 

\subsubsection*{Conflicts of interest}
 The author declares no conflicts of interest regarding the publication of this paper.

\end{document}